\documentclass[]{article}
\usepackage{graphicx} 
\usepackage{placeins}
\usepackage{float}

\title{A High Accuracy Image Hashing and Random Forest Classifier for Crack Detection in Concrete Surface Images}
\author{Diego Frias and José Hidalgo}

\begin{document}

\maketitle

\begin{abstract}
	Automatic detection of cracks in concrete surfaces based on image processing is a clear trend in modern civil engineering applications. Most infrastructure is made of concrete and cracks reveal degradation of the structural integrity of the facilities, which can lead to extreme structural failures. There are many approaches to overcome the difficulties in image-based crack detection, ranging from the pre-processing of the input image to the proper adjustment of efficient classifiers, passing through the essential feature selection step. This paper is related to the process of constructing features from images to allow a classifier to find the boundaries between images with and without cracks. The most common approaches to feature extraction are the convolutional techniques to extract relevant positional information from images and the filters for edge detection or background removal. Here we apply hashing techniques for the first time used for features extraction in this problem. The study of the classification capacity of hashes is carried out by comparing 5 different hash algorithms, 2 of which are based on wavelets. The effect of applying the z-transform on the images before calculating the hashes was also studied, which totals the study of 10 new features for this problem. A  comparative study of 17 different algorithms from the scikit-learn library was carried out. The results show that 9 of the 10 features are relevant to the problem, as well as that the accuracy of the classifiers varied between 0.697 for the Naive-Bayes Gaussian classifier and 0.99 for the Random Forest (RF) classifier. The feature extraction algorithm developed in this work and the RF classifier algorithm is suitable for embedded applications, for example in inspection drones, as long as they are highly accurate and computationally light, both in terms of memory and processing time.
	
\end{abstract}


\section{Introduction}
Concrete is an extremely important material in civil construction, being one of the main components of a variety of structures of great economic and social value such as bridges, buildings, dams, stadiums and so on. Like many materials, concrete is subject to physical changes caused by various factors such as increased strength on the structure, wear caused by flooding, corrosion of internal metal structures and others. When a concrete structure begins to fail caused by any of these factors, one of the most notable symptoms is the appearance of cracks that can usually be seen from the surfaces of the structures.
Thus, the detection of these cracks constitutes an important activity in the preventive maintenance of concrete structures.

As stated in Park et al. (2006), the ways to analyze the health of a concrete structure are relatively underdeveloped compared to metallic structures and other composite materials. Some approaches to crack detection in concrete have been developed over the years, the main ones are: detection from acoustic emission (OHNO; OHTSU, 2010) and detection based on impedance (PARK et al., 2006), among others.

In recent years the field of artificial intelligence has experienced great development, mainly due to the lower cost and better performance of hardware such as CPUs, GPUs and RAM memories, factors that allowed a computational model developed in the 1980s to gain much popularity: the Neural Network , which has performed very well in image classification tasks in different applications, including the detection of concrete cracks (KRIZHEVSKY; SUTSKEVER; HINTON, 2012) (GOODFELLOW et al., 2016) (YAO G. et al., 2018) (SHENGYUAN L. and XUEFENG Z., 2019) and (LIU K. et al., 2019).

However, despite the excellent performance of neural networks in most applications (NEUROHIVE, 2018) (RUSSAKOVSKY, O. et al, 2015), they have the disadvantage of having a higher computational cost (memory and computing power) than non-neural models. Therefore, when choosing a model to be embeded in a mobile inspection device, such as a drone, neural models do not have priority (REZAIE F, ASGARINEJAD M, 2020) (FEROZ, S. and ABU DABOUS, S, 2021) and (LIU K. et al., 2019).

For this reason, it is important to identify which other non-neural classification methods are feasible to use for embedded applications, seeking to identify the one with the best cost-benefit ratio. At the same time, it is important to explore other alternatives for extracting features from the images of the concrete surfaces to be processed, which allows to reduce the number of them, which positively impacts the volume of data to be processed and accordingly the processing time, which is relevant for real-time applications.

An extensive review of literature showed that a wide variety of features have been proposed and validated for this application, highlighting those extracted by hierarchical spatial convolution (used in CNN convolutional neural networks) (YAO G. et al., 2018) (SHENGYUAN L. and XUEFENG Z., 2019), and those extracted with computer vision techniques (equalization, background removal, location of edges, and segmentation) (YAMAGUSHI T. et al.,2008) (LIANG S. et al., 2018).

In this article, we address the problem of building features never before applied to detect cracks in concrete surfaces and finding the best non-neural model for these features. We use five different image hashing algorithms to assign a unique hash value to an image, often used as a "digital fingerprint" in image retrieval systems. One of the perceived advantages of hashing is that the hash number of an image is practically invariant in the face of format or resolution changes, or even if there is a slight corruption, perhaps due to previous compression.

Our initial hypothesis was that a right mix of digital fingerprints of the images should carry a sufficient amount of information, that would allow training highly accurate and precise non-neural classifiers.

All features in image based processing constructing features that consists in extracting, from the matrix of pixels that make up the images, a set of input variables that contain enough information to allow the classifier algorithm to define the boundaries between two classes: images containing crack and images not containing crack. The most common approaches to feature extraction are the use of convolutional techniques to extract relevant positional information from images or filters for edge detection/background extraction. Here we use a technique that reduces all information in images to a single 64-bit integer, called a hash. According to the authors' knowledge, this is the first time that the image hashing technique is used to generate features in the problem of detecting cracks in concrete. The research of the classification capacity of hashs is carried out by comparing 5 different hash algorithms, 2 of which are based on wavelets. The effect of applying the z-transform on the images before calculating the hashs was also studied, which totals the study of 10 new features for this problem. In relation to the classifiers, the study of 17 different algorithms from the scikit-learn library was carried out. 

A sequential feature selecion procedure showed that 9 of the 10 features are relevant to the problem, as well as that the best classifiers are of ensemble type, being the Random Forest (RF) classifier the best in our image processing and classification framework. 

The results confirmed our hypothesis, being possible to build a very accurate and precise non-neural classifier (accuracy = precision = recall = 0.99) using a set of 9 image hashing features.

The developed model, composed of a feature extractor and a RF classifier, can be used in embedded applications, such as inspection drones, as it uses a small data volume and short processing time.

This article is structured as follows: In the section \ref{sec:data_algo}, we describe the hashing algorithms, the cross-validation procedure, the source, the size and composition of the datasets, the classifier models and the metrics of performance scores used. The \ref{sec:results} section discusses the model selection and the best results of the final model test. Finally, in the \ref{sec:conclusions} section the final remarks are made.

\section{Data and Algorithms}\label{sec:data_algo}

Images were downloaded from the Concrete Crack for Image Classification dataset (ÖZGENEL, 2018), which contains 20,000 images of concrete surfaces with cracks and 20,000 images without cracks. All images have dimensions of $227\times 227$ pixels. All images were randomly generated from $458$ high-resolution images ($4032\times3024$ pixels), and the images show different lighting. 

We used the Mean, Difference, Perceptual and Wavelet hashing algorithms for feature generation. All of them initially generate a grayscale image from the input image, which is resized to a smaller size image, which is later binarized.

The Average hash (ahash) averages the pixels and binarizes the reduced grayscale image $8 \times 8 $ based on the average. If the pixel is larger than average, it assigns 1, otherwise 0. The binary image is then transformed into the 64-bit hash integer.

The Difference hash (dhash) scales the grayscale image to $8 \times 9$ and compares the first 8 pixels of each row with the neighbor on the right. If it is less than the neighbor it assigns 0, otherwise 1. In this way, a binary array $8 \times 8$ is obtained and transformed into a 64-bit hash.

The Perceptual hash (phash) algorithm reduces the grayscale image to $8f\times 8f$ pixels, where $f$ is a scaling factor. Next, a discrete cosine transform, 
first per row and then per column, is applied, resulting in an image where 
the pixels with the highest frequencies are located in the $8\times8$ upper left corner. Then the corner image is cropped and the median of the gray values is calculated. The median is used to binarize the image obtained, similar to 
what is done in the Average hash method.

The Wavelet hash (whash) algorithm applies a two-dimensional wavelet transformation to the $8 \ times8 $ reduced grayscale image. Then, analogously to the perceptual hash algorithm, each pixel is compared to the median and the hash is calculated. Two mother wavelet functions were used: Daubechies 4 and Haar (Daubechies 1), as suggested in the literature for texture recognition. For this reason, we denote whash-haar and whash-db4 these two image hashing methods.

Furthermore, in order to expand our feature set from 5 to 10, we introduced a scale reduction algorithm based on the 2D Z transform (SENGUPTA M. and MANDAL JK., 2013), which replaced the usual scale reduction algorithm based on local averages. Finally, our feature set consists of:
\begin{enumerate}
	\item ahash, 
	\item dhash, 
	\item phash, 
	\item whash-haar, 
	\item whash-db4, 
	\item z-transf-ahash, 
	\item z-transf-dhash, 
	\item z-transf-phash, 
	\item z-transf-whash-haar, and 
	\item z-transf-whash-db4. 
\end{enumerate}

It should be clarified that 64-bit integer hashs were converted to float before being used to train classifiers.

Once the feature set was established, the set of non-neural classifiers was defined as:
\begin{enumerate}
	\item Logistic Regression Classifier
	\item Ridge Classifier
	\item SGD Classifier:, cv=0.839
	\item Passive Aggressive Classifier
	\item KNeighbors Classifier
	\item Decisio Tree Classifier
	\item Extra-Tree Classifier
	\item Linear Support Vector Classifier 
	\item Support Vector Classifier
	\item Gaussian Naive-Bayes Classifier
	\item AdaBoost Classifier
	\item Bagging Classifier
	\item Random Forest Classifier
	\item Extra-Trees Classifier
	\item Gradient Boosting Classifier
	\item Linear Discriminant Analysis Classifier
	\item Quadratic Discriminant Analysis Classifier
\end{enumerate} 

Classifiers were imported from the scikit-learn 0.24.2 python library (PEDREGOSA F. et al., 2011).

We use the K-fold cross validation procedure that randomly divides the training dataset into K parts of the same size, using each part to evaluate the model trained with the other K-1 parts, so that after K rounds of training and test, an average performance metric is calculated for each model. Our training dataset consisted of 20,000 images (50\% of all available dataset), with 10,000 images with cracks and 10,000 without cracks. The performance metric chosen was accuracy and K = 10. Therefore, the 17 classifiers were trained 10 times with non-identical sets of 18,000 images each and 10 times tested with non-identical sets of 2000 images.

The best performing model was then trained with 70 \% of the dataset used for cross-validation, that is, with a randomly sampled set of 14,000 images, and tested with a set of 20,000 images not used for cross-validation, ie, the second half of the original dataset, also containing 10,000 cracked and 10,000 uncracked.

To assess the performance of the classifier with the best score in the large test sample, we used the standard set of supervised score metrics: Accuracy, Precision, Recall, and F1-score, plus the area under the receiver operating curve, (AUC-ROC).

\section{Results}\label{sec:results}

\begin{figure}[H]
	\centering
	\includegraphics[width=0.9\linewidth]{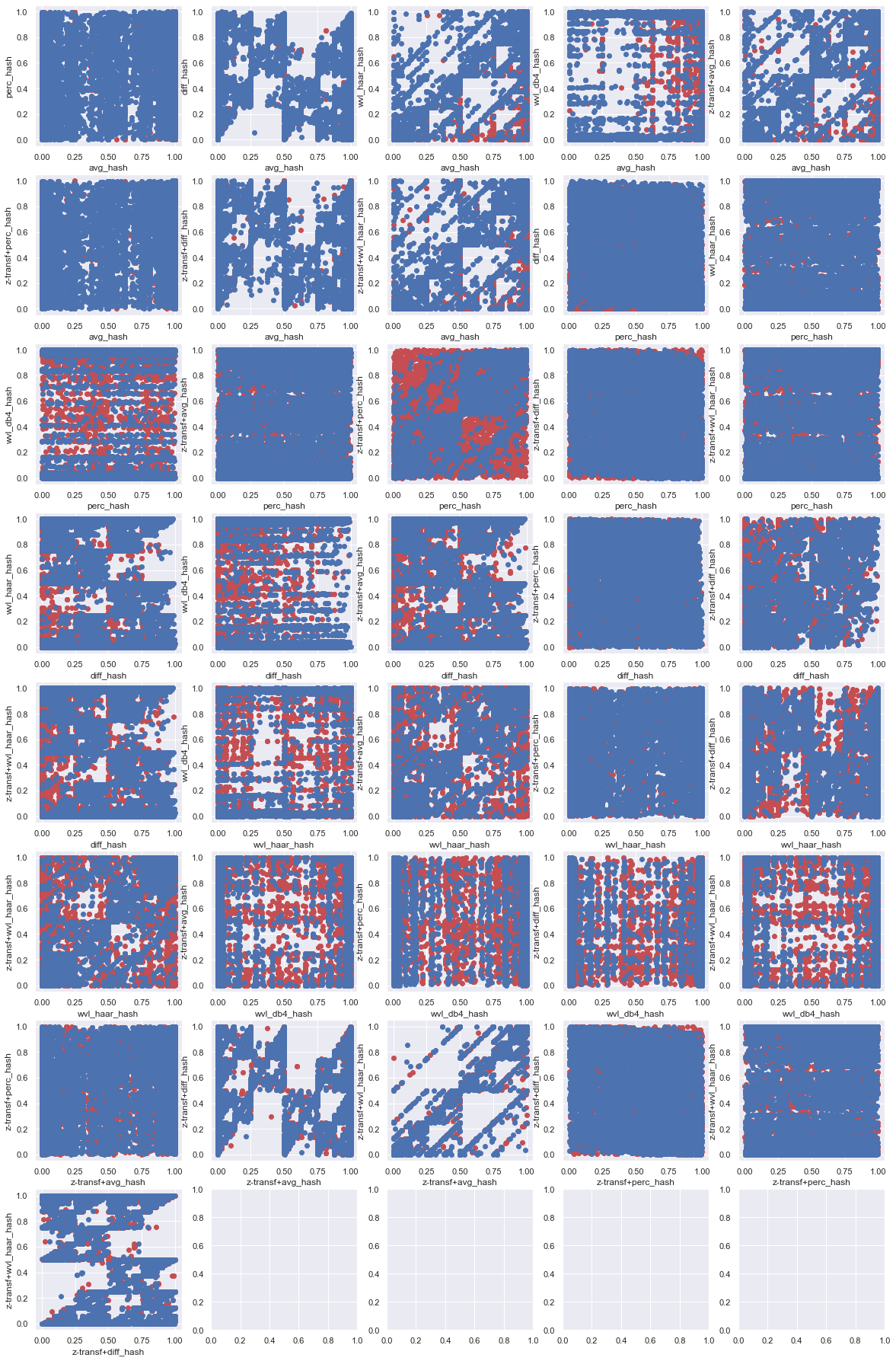}
	\caption{Scatter plot of the train samples for each pair of features. Red dots are images with cracks and blue dots without cracks.}
	\label{fig:colagefeaturepairs}
\end{figure}

In figure \ref{fig:colagefeaturepairs} we show the distribution of images by class (cracked=red or not cracked=blue) for each pair of hash-based features. Notice that there is not any pair of features where the two classes form separated clusters, which indicates that the classification problme is quite hard with this set of features.
\FloatBarrier

\begin{figure}[H]
	\centering
	\includegraphics[width=0.9\linewidth]{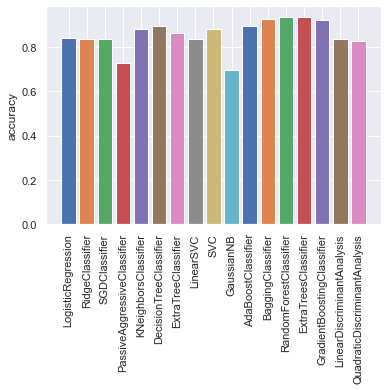}
	\caption{Performance of classifiers with the new feature set on 20,000 images during 10-fold cross-validation}
	\label{fig:classifieraccuracies}
\end{figure}

In figure \ref{fig:classifieraccuracies} and table \ref{tab:classifiers_accuracy} we show the performance (average accuracy in cross validation) of the 17 studied classifiers. The best classifier was Forest Random (accuracy = 0.940), closely followed by the Extra-Tree classifier (accuracy = 0.936). 

It is important to note that the top five classifiers have an "ensemble" structure consisting of a large number of independently trained weak (simple) learners whose predictions are combined to build a strong classifier (SAGI, O and ROKACH, L., 2018) .

\begin{table}[H]
	\centering
	\caption{Performance of classifiers with the new feature set on 20,000 images during 10-fold cross-validation, ordered from best to worst.}
	\label{tab:classifiers_accuracy}
\begin{tabular}{|c|c|}
	\hline
	Classifier & Average Accuracy in 10-fold validation \\
	\hline
	Random Forest Classifier&0.940 \\
	Extra-Trees Classifier&0.936 \\
	Bagging Classifier&0.928 \\
	Gradient Boosting Classifier&0.923 \\
	AdaBoost Classifier&0.899 \\
	Decision Tree Classifier&0.896 \\
	KNeighbors Classifier&0.883 \\
	Support Vector Classifier&0.883 \\
	Extra-Tree Classifier&0.865 \\
	Logistic Regression & 0.840 \\
	Linear Support Vector Classifier&0.839 \\
	Linear Discriminant Analysis Classifier &0.838 \\
	Ridge Classifier&0.838 \\
	SGD Classifier&0.836 \\
	Quadratic Discriminant Analysis Classifier &0.829 \\
	Passive Aggressive Classifier&0.728 \\
	Gaussian Naive-Bayes Classifier&0.697 \\
	\hline
\end{tabular}
\end{table}

\begin{figure}[H]
	\centering
	\includegraphics[width=0.7\linewidth]{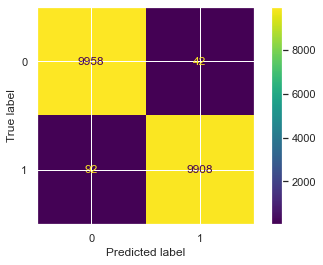}
	\caption{Confusion Matrix of Random Forest Classifier with a perfectly balanced test set of 20,000 images.}
	\label{fig:confusionmatrixrfconcrete}
\end{figure}

In figure \ref{fig:confusionmatrixrfconcrete} we show the confusion matrix of Random Forest Classifier with a perfectly balanced test set of 20,000 images, and in table

\begin{table}
	\centering
	\caption{Performance metrics of the Random Forest classifier in a test dataset with 20,000 images.}
	\label{tab:RFreport}
\begin{tabular}{|c|c|c|c|c|}
	\hline
	&  precision  &   recall & f1-score  & support\\
    \hline
	unfratured &      0.99  &    1.00 &     0.99  &    10,000\\
	fractured  &     1.00   &   0.99  &    0.99  &    10,000\\
	accuracy   &            &        &    0.99  &    20,000\\
	macro avg   &    0.99   &   0.99   &   0.99   &   20,000\\
	weighted avg   &    0.99  &    0.99   &   0.99  &    20,000\\
	\hline
\end{tabular}
\end{table}

Random Forest Classifier's impressive performance with a set of just 10 features (floats) extracted from the images using simple hashing techniques, confirmed our initial hypothesis that a right mix of "digital fingerprints" of the images should carry a sufficient amount of information, that would allow training highly accurate and precise non-neural classifiers.

\begin{figure}[H]
	\centering
	\includegraphics[width=0.8\linewidth]{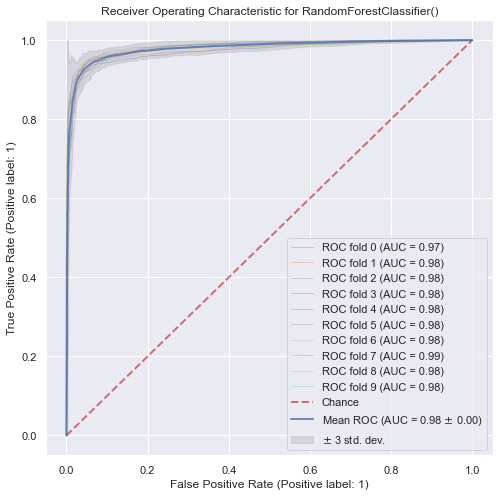}
	\caption{Receiver Operating Characteristic curve and 10-fold cross validation area under the curve (AUC) for the Random Forest Classifier with the new feature set.}
	\label{fig:aucrocrf}
\end{figure}

Exceptionally, all metrics measured during the classification of a large test dataset, 43\% larger than the training dataset, ranged between 0.99 and 1.00. Even more surprising is the near-ideal behavior of the ROC curve (shown in figure \ref{fig:aucrocrf}), which summarizes the trade-off between the true-positive rate and the false-positive rate for different probability thresholds, used to define the prediction of the classifier. The ideal classifier has ROC-AUC = 1.0 and our classifier achieves ROC-AUC = 0.98, with negligible dispersion.

\section{Concluding Remarks}\label{sec:conclusions}

In this work, the use of image hashing as features for the classification of concrete images with and without cracks was discussed for the first time. This is particularly important because most civil infrastructure is made of concrete and cracks reveal the degradation of the structural integrity of the facilities, which can lead to extreme structural failures. For this reason, automated detection of cracks in concrete surfaces based on image processing is a clear trend in modern civil engineering applications. Several projects develop land, sea and air drones for remote inspection of structures. These devices require agile and lightweight algorithms (low memory consumption) to be embedded in order to perform in situ and real-time evaluations.

Deep learning approaches based on different variants of Convolutional Neural Networks have been used successfully for this purpose, but due to the large volume of data that needs to be processed, these neural models are not the best options to be embedded in today's drones.

Despite having found in the literature non-neural models for classification of images with and without cracks, based on features extracted with computer vision techniques, we realize that: (1) the maximum reported score (accuracy) varies between 0.87 and 0.93, (2) the test datasets in most cases are not very representative of reality and, (3) a reliable statistical validation procedure of the classifiers, such as the K-fold cross-validation used in this work, was not used.

For this reason, we carried out a study using a perfectly balanced dataset, with 40,000 images, subject to natural disturbances (resolution, luminosity, geometric distortion, post-compression and decompression data corruption, etc.) and implemented a statistically based validation process of the performance of the models.

Another differential of our approach is the use of image hashing features, generally used for storing and retrieving images in databases, but never used before for this problem.

After testing 17 non-neural classifiers we find that a Random Forest classifier trained with an small set of features, consisting of 10 different 64-bit hashes taken from concrete image of $227\times227$ pixels, achieves very high scores,  with accuracy, precision, recall and F1-score ranging between 0.99 and 1.0 and an AUC-ROC = 0.98. In other words, we show the extremely high capacity of a non-neural model to classify crack-free and crack-free concrete images using few image hashing features.

\section*{References}

\begin{enumerate} 
	
	\item FEROZ, S.; ABU DABOUS, S. UAV-Based Remote Sensing Applications for Bridge Condition Assessment. Remote Sens. 2021, 13, 1809. https://doi.org/10.3390/rs13091809.
	
	\item GOODFELLOW, I. et al. Deep learning. [S.l.]: MIT press Cambridge, 2016. v. 1. 
	
	\item KRIZHEVSKY, A.; SUTSKEVER, I.; HINTON, G. E. Imagenet classification with
	deep convolutional neural networks. In: Advances in Neural Information Processing Systems 25: 26th Annual Conference on Neural Information Processing Systems	2012. p. 1106–1114. Disponível em: http://papers.nips.cc/paper/4824-imagenet-classification-with-deep-convolutional-neural-networks. 
	
	\item LIANG S., JIANCHUN X. AND XUN Z., "An Extraction and Classification	Algorithm for Concrete Cracks Based on Machine Vision," in IEEE	Access, vol. 6, pp. 45051-45061, 2018, doi:	10.1109/ACCESS.2018.2856806.
		
	\item LIU K., HAN X. AND CHEN BM., "Deep Learning Based Automatic Crack Detection and Segmentation for Unmanned Aerial Vehicle Inspections," 2019 IEEE International Conference on Robotics and Biomimetics (ROBIO), Dali, China, 2019, pp. 381-387, doi:	10.1109/ROBIO49542.2019.8961534.
	
	\item NEUROHIVE. VGG16 – Convolutional Network for Classification and Detection. 2018. Disponível em: https://neurohive.io/en/popular-networks/vgg16/. 
	
	\item OHNO, K.; OHTSU, M. Crack classification in concrete based on acoustic emission. Construction and Building Materials, Elsevier, v. 24, n. 12, p. 2339–2346, 2010. 
	
	\item ÖZGENEL, Ç. F. Concrete crack images for classification. Mendeley Data, v1 http://dx.doi. org/10.17632/5y9wdsg2zt, v. 1, 2018. 
	
	\item PARK, S. et al. Multiple crack detection of concrete structures using impedance-based structural health monitoring techniques. Experimental Mechanics, Springer, v. 46, n. 5, p. 609–618, 2006. 
	
	\item PEDREGOSA, F. et al. Scikit-learn: Machine learning in Python. Journal of Machine Learning Research, v. 12, p. 2825–2830, 2011. 
	
	\item REZAIE F, ASGARINEJAD M. A New Method Of Studying Of Different Types Of Fractures And Cracks In Concrete By Image Processing. Journal of Multidisciplinary Engineering Science and Technology (JMEST). ISSN: 2458-9403 Vol. 7 Issue 2, February - 2020.
	
	\item RUSSAKOVSKY, O. et al. ImageNet Large Scale Visual Recognition Challenge.
	International Journal of Computer Vision (IJCV), v. 115, n. 3, p. 211–252, 2015.
	
	\item SAGI, O, ROKACH, L. Ensemble learning: A survey. WIREs Data Mining Knowl Discov. 2018; 8:e1249. https://doi.org/10.1002/widm.1249
	
	\item SENGUPTA M., MANDAL JK. Image Coding Through Z-transform With Low Energy and Bandwidth (IZEB). Jan Zizka (Eds) : CCSIT, SIPP, AISC, PDCTA - 2013
	pp. 225–232, 2013. © CS \& IT-CSCP 2013 DOI : 10.5121/csit.2013.3625
	 	
	\item SHENGYUAN L., XUEFENG Z., "Image-Based Concrete Crack Detection
	Using Convolutional Neural Network and Exhaustive Search Technique", Advances in Civil Engineering, vol. 2019, Article ID 6520620, 12 pages, 2019.
	
	\item YAMAGUCHI T., NAKAMURA S. AND HASHIMOTO S., "An efficient crack detection method using percolation-based image processing," 2008 3rd IEEE Conference on Industrial Electronics and Applications, Singapore, 2008, pp. 1875-1880, doi:10.1109/ICIEA.2008.4582845.
	
	\item YAO G., WEI W., QIAN J. and WU Z., "Crack Detection Of Concrete
	Surface Based On Convolutional Neural Networks," 2018
	International Conference on Machine Learning and Cybernetics
	(ICMLC), Chengdu, 2018, pp. 246-250, doi:
	10.1109/ICMLC.2018.8527035. p. 29–39.

\end{enumerate}

\end{document}